\documentclass[10pt,aps,twocolumn,pra,superscriptaddress]{revtex4-1}
\usepackage{amsfonts}

\usepackage{amsmath}
\usepackage{graphicx,color}
\usepackage{epstopdf}
\usepackage{epsfig}
\usepackage{mathrsfs}
\usepackage[breaklinks=true]{hyperref}

\newcommand{\beq}{\begin{equation}}
\newcommand{\eeq}{\end{equation}}
\newcommand{\bqa}{\begin{eqnarray}}
\newcommand{\eqa}{\end{eqnarray}}

\definecolor{green}{rgb}{0.00,0.50,0.00}

\newcommand{\Eye}{\hbox{1\kern-4truept 1}}

\begin{document}

\title{The Collapse Before a Quantum Jump Transition}

\author{John E.~Gough} \email{jug@aber.ac.uk}
   \affiliation{Aberystwyth University, SY23 3BZ, Wales, United Kingdom}
\date{\today}

\begin{abstract}
We may infer a transition $|n \rangle \to |m \rangle$ between energy eigenstates of an open quantum system by observing the emission of a photon of Bohr frequency $\omega_{mn} = (E_n-E_m) / \hbar$. In addition to the \lq\lq collapses\rq\rq   to the state $|m\rangle$, the measurement must also have brought into existence the pre-measurement state $|n \rangle$. As quantum trajectories are based on past observations, the condition state will jump to $| m \rangle$, but the state $|n\rangle$ does not feature in any essential way. We resolve this paradox by looking at quantum smoothing and derive the time-symmetric model for quantum jumps.
\end{abstract}

\maketitle

\affiliation{Aberystwyth University, SY23 3BZ, Wales, United Kingdom}
 
\bigskip

\section{Introduction}
In the Copenhagen interpretation, an observable $A$ does not possess an actual value until we measure it, in which case we observe an eigenvalues $a$ and the state collapses into the eigenstate $| a\rangle $. This assumes a direct measurement. In practice, we perform indirect measurements and make inferences on what the state must be.

Consider, an atomic electron with a complete orthonormal basis $\left\{ |n\rangle :n=0,1,2,\cdots \right\} $ of energy eigenstates with corresponding energy eigenvalues $ \{
E_{n}: n=0,1,,2,\cdots  \} $; when the atom emits a photon its frequency must be one of the Bohr frequencies $\omega _{mn}=E_{n}-E_{m}$. Furthermore let us assume that the set of Bohr frequencies is non-degeneracy so that is we observe a photon with the frequency $\omega_{mn}$ then we know that the electron has undergone a transition $|n\rangle \rightarrow |m\rangle $. 
\begin{figure}[h]
	\centering
		\includegraphics[width=0.30\textwidth]{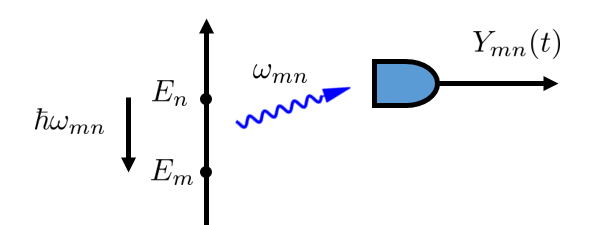}
	\caption{(color online) Detecting a transiotion $|n \rangle \to | m \rangle$ with emitted photon.}
	\label{fig:Bohr_transition}
\end{figure}

Conventional theory tell is that the state collapses to $|m \rangle$ at the time of measurement: but in fact we learn more! We are not measuring the Hamiltonian $H$ of the electron, but instead measuring a transition between its eigenstates, and so also infer that the state of the system immediately before measurement was $|n \rangle$: despite making no assumptions on the initial state!

There has been much interest in using the full data recorded through continual quantum measurement of an open system to estimate, for instance, indirect measurement made during the monitoring period. This had lead to a time-symmetric theory.
interventions in quantum measurement \cite{Theory_Past}, as well as recent experimental tests \cite{Minev18}.
In \S II we outline the theory, and in \S III derive the time-symmetric form. Here we must derive the result for photon counting as opposed to homodyne measurement of quadratures from previous papers.

\section{Estimating Bohr Transitions}

Let us now describe the model. We assume, for simplicity, that we have an $N$-level system with Hamiltonian
$ H_{\text{sys}} =
 \sum_{n=0}^{N-1} E_{n}|n\rangle \langle n|$
with energy eigenvalues $E_{0}<E_{1}<\cdots <E_{N-1}$. The set, $\mathscr{F}$, of Bohr frequencies is then the collection $\omega _{mn}=E_{n}-E_{m}$. The positive Bohr frequencies are then $\omega _{mn}$
with $m<n$ and we assume that they are non-degenerate.

The system couples to a bath (the quantum electromagnetic field) and the evolution is described by the quantum stochastic differential equation (QSDE)
\begin{eqnarray}
dU\left( t\right)  = \bigg\{ \sum_{m<n} L_{mn} \otimes dB_{mn}\left( t\right) ^{\ast }-\text{H.c.}  +K\otimes dt \bigg\} U\left( t\right) \nonumber \\
\label{eq:U}
\end{eqnarray}
where the $B_{mn}\left( t\right) ,B_{mn}\left( t\right) ^{\ast }$ are independent pairs of annihilation and creation processes \cite{HP84,GarCol85} for the bath and we have the collapse operators and non-hermitean damping operators
(This is derived in the Appendix.)
\begin{eqnarray}
L_{mn} = \sqrt{\gamma _{mn}}|m\rangle \langle
n| , \quad
K=-\frac{1}{2}\sum_{m<n}\gamma _{mn}|n\rangle \langle n|.
\label{eq:K}
\end{eqnarray}

\subsection{The Unconditioned Evolution}

Let us take the initial state of the system to be $\rho _{0}$ and the state of the bath to be the vacuum $|\Omega \rangle $. For a given system operator $X$ we obtain its expectation at time $t$ as
\begin{eqnarray*}
\left\langle X\right\rangle _{t} &=&\text{tr}\left\{ \rho _{0}\otimes
|\Omega \rangle \langle \Omega |\,U\left( t\right) ^{\ast }\left[ X\otimes
\Eye_{\text{bath}}\right] U\left( t\right) \right\},   
\end{eqnarray*}
or $\left\langle X\right\rangle _{t} = \text{tr}\left\{ \rho _{0}\Phi _{t}\left( X\right) \right\} $ where $\Phi _{t}$ is the quantum Markov semi-group associate with (\ref{eq:U}) with Lindbladian 
\begin{eqnarray*}
\mathcal{L}\left( X\right) =\sum_{m<n}\gamma _{mn}\left\{ \langle
m|X|m\rangle \,|n\rangle \langle n|-\frac{1}{2}\left[ |n\rangle \langle n|,X%
\right] _{+}\right\} . \nonumber \\
\,
\label{eq:Lindbladian}
\end{eqnarray*}

The system state evolves according to $tr\left\{ \rho _{t}X\right\} \equiv tr\left\{ \rho _{0}\Phi _{t}\left( X\right) \right\} $ and one readily deduces the master equation $\frac{d\rho _{t}}{dt} = \mathcal{L}^{\ast }\left( \rho _{t}\right)$ where the Liouvillian 
(adjoint of $\mathcal{L}$) is
\begin{eqnarray*}
\mathcal{L}^{\ast }\left( \rho  \right) = \sum_{m<n}\gamma _{mn}\left\{ \langle n|\rho |n\rangle \,|m\rangle \langle
m|-\frac{1}{2}\left[ |n\rangle \langle n|,\rho \right] _{+}\right\} .
\end{eqnarray*}

For instance, let us take the projection $P_{n}=|n\rangle \langle n|$ and set $p_{n}\left( t\right) =\left\langle P_{n}\right\rangle _{t}$, then we obtain
\begin{eqnarray}
\frac{d}{dt}p_{r}\left( t\right) =\sum_{n>r}\gamma _{rn}p_{n}\left( t\right)
-\sum_{m<r}\gamma _{mr}p_{r}\left( t\right) .
\label{eq:p_ODE}
\end{eqnarray}
Here we see that $\sum_{n}\dot{p}_{n}\left( t\right) \equiv 0$, and the unique asymptotic limit is $p_{0}\left( t\right) \rightarrow 1$ with all other probabilities tending to zero. In other words, the electron eventually decays to the ground state.

\subsection{Monitoring Transitions}

We now wish to describe the measurement of a photon of frequency $\omega_{mn}$. Let us introduce the bath variable
\begin{eqnarray}
Z_{mn}\left( t\right) =\int_{0}^{t}b_{mn}\left( s\right) ^{\ast
}b_{mn}\left( s\right) ds
\end{eqnarray}
corresponding to the number of input photons resonant with the Bohr frequency $\omega _{mn}$ during the time interval 0 to $t$. Of, course, what we want to measure is the number in the output channel and this is described by the observable
\begin{eqnarray}
Y_{nm}\left( t\right) =U\left( t\right) ^{\ast }\left[ \Eye_{\text{sys}}\otimes
Z_{mn}\left( t\right) \right] U\left( t\right) .
\label{eq:Y}
\end{eqnarray}

The family $\left\{ Y_{mn}\left( t\right) :t\geq 0,m<n\right\} $ is self-commuting and will give the set of measured observables. Formally we have the output fields
\begin{eqnarray*}
b_{mn}^{\text{(out)} }(t)  = b_{mn} (t) + \sqrt{\gamma_{mn}} U\left( t\right) ^{\ast }\left[ |m \rangle \langle n|
 \otimes \Eye_{\text{bath}}
 \right] U\left( t\right)  ,
\end{eqnarray*}
and $dY_{mn} (t) \equiv b_{mn}^{\text{(out)}} (t) ^\ast b_{mn}^{\text{(out)}} (t)  \, dt$.

Noting that the future-pointing (Ito) increments $dZ_{mn},$ $ dB_{mn} (t)$, and $ dB_{mn}(t)^\ast$ vanish in the vacuum state for the bath, we see that average increment of the observed output is
\begin{eqnarray}
\text{tr}\left\{ \rho _{0}\otimes
|\Omega \rangle \langle \Omega | \, dY_{mn} (t) \right\} =
\gamma_{mn} p_n (t) \, dt.
\label{eq:Y_rate}
\end{eqnarray}
In particular, $Y_{mn} (t)$ is a time-inhomogeneous Poisson process whose rate is given by (\ref{eq:Y_rate}).

The goal of quantum filtering is to give the least-squares estimate for any observable $X$ of the system at time $t$ given the observations up to that time. We denote this as $\pi _{t}\left( X\right) $ and it will satisfy the stochastic differential equation \cite{Belavkin}
\begin{eqnarray}
d\pi _{t}\left( X\right) &=&\pi _{t}\left( \mathcal{L} X\right) dt \label{eq:filter} \\
&&+\sum_{m<n}\bigg\{
\langle m|X|m\rangle -\pi _{t}\left( X\right) \bigg\} \,dI_{mn}\left(
t\right) 
\nonumber
\end{eqnarray}
where the $I_{mn}\left( t\right) $ are defined by
\begin{eqnarray}
I_{mn}\left( t\right) =Y_{mn}\left( t\right) -\gamma _{mn}\int_{0}^{t}\pi
_{s}\big( |n\rangle \langle n|\big) ds.
\label{eq:innovations}
\end{eqnarray}
The stochastic process $I_{mn}$ is martingale \cite{martingale} with respect to the observations, and in the present case it is a time-inhomogeneous, compensated Poisson process.

The result may be alternatively stated as a stochastic master equation (SME) \cite{CarBook93,DCM}. The state $\hat{\rho}_{t}$ of the system at time $t$ conditional on the observations up to that
time is defined by $\pi _{t}\left( X\right) =$tr$\left\{ \hat{\rho}_{t}X\right\} $, and satisfies
\begin{eqnarray}
d\hat{\rho}_{t}=\mathcal{L}^{\ast }\left( \hat{\rho}_{t}\right)
dt+\sum_{m<n}\bigg\{ |m\rangle \langle m|-\hat{\rho}_{t}\bigg\}
\,dI_{mn}\left( t\right) .
\label{eq:SME}
\end{eqnarray}

The SDE for  $\rho_t $ is nonlinear, but we may write conditional averages as $\pi_t (X) \equiv \sigma_t (X) / \sigma_t (\Eye)$ where $\sigma_t (X)$ satisfies a linear SDE, known as the Belavkin-Zakai equation. This has been calculated for jump processes, \cite{BvH_reference} \S 7.3, and takes the form
\begin{eqnarray}
d \sigma_t (X) &=& \sigma_t ( \mathcal{L} X) dt 
\label{eq:BZ_jump} \\
&+& \sum_{m<n} \sigma_t (L_{mn}^\ast X L_{mn} -X) [ dY_{mn} (t) -dt ]. \nonumber
\end{eqnarray}

Let us remark that, over periods where no photon counts are made, we have $dY_{mn}\left( t\right) \equiv 0$, and so $\hat{\rho}$ evolves according to the deterministic ODE
\begin{eqnarray}
\frac{d\hat{\rho}_{t} }{dt} \bigg|_{\text{no count}}  &=&\mathcal{L}^{\ast }\left( \hat{\rho}_{t}\right) \label{eq:phat_no_count}\\
&-& \sum_{m<n}\gamma _{mn}\bigg\{ |m\rangle \langle m|-\hat{\rho}_{t}\bigg\}
\langle n|\hat{\rho}_{t}|n\rangle . 
\nonumber
\end{eqnarray}
This implies that, between counts, the probabilities $\hat{p}_{r}\left(
t\right) =\langle r|\hat{\rho}_{t}|r\rangle =\pi _{t}\left( |r\rangle
\langle r|\right) $ evolve according to
\begin{eqnarray}
\frac{d\hat{p}_{r}\left( t\right)}{dt} \bigg|_{\text{no count}} =-\left( \sum_{m<r}\gamma
_{mr}-\sum_{m<n}\gamma _{mn}\hat{p}_{n}\left( t\right) \right) \hat{p}%
_{r}\left( t\right) . \nonumber \\
\,
\label{eq:phat_ODE}
\end{eqnarray}
The nonlinearity is a feature of the continual measurement back action. More exactly, (\ref{eq:phat_ODE}) gives the evolution of the probabilities to be in state $| r \rangle$ conditional that we observe \textit{no photon counts}: as opposed to (\ref{eq:p_ODE}) which gives the evolution of the probability that we ignore any counts.

We note that (\ref{eq:phat_ODE}) may be rewritten as
\begin{eqnarray}
\frac{d}{dt}\hat{p}_{r}\left( t\right) =-\bigg( \Gamma_r -A ( \mathbf{p}(t) )  \bigg) \hat{p}_{r}\left( t\right) .
\label{eq:phat_ODE1}
\end{eqnarray}
where $\Gamma_r = \sum_{m<r}\gamma_{mr}$ and $A (\mathbf{p}) =\sum_{m<n}\gamma _{mn}\hat{p}_{n}$. We have $\Gamma_0 = 0$ but $\Gamma_r >0$ for $r \geq 1$.

The situation where the system is in a given state $|k \rangle$ corresponds to $ \mathbf{p} = \delta_k$, that is, $p_k = 1$ and all other $p_r =0$. We have $A ( \delta_k ) \equiv \Gamma_k$, and we see readily that (\ref{eq:phat_ODE1}) has the equilibrium points $\mathbf{p} = \delta_k$, for each $k$. However, only the ground state $k=0$ is asymptotically stable. Specifically, if we linearize around $\delta_k$ for $k>0$ we find $\frac{d}{dt} \hat{p}_r \approx (\Gamma_r - \Gamma_k ) \hat{p}_r$ for $r \neq k$ and which imply exponential growth for $r <k$. 
 We also see that
\begin{eqnarray}
\frac{d}{dt}\sum_r \hat{p}_{r}\left( t\right) =A ( \mathbf{p}(t) ) \bigg( \sum_r \hat{p}_{r}\left( t\right) - 1  \bigg)   ,
\label{eq:phat_ODE2}
\end{eqnarray}
implying that $\sum_r \hat{p}_{r}\left( t\right) =1$ throughout the evolution.

\bigskip

\section{Observing a Transition Event}

Having set up the model, we now look at what typically happens. From an experimental point of view, we are monitoring the output light field and watching for photons resonant with one of the Bohr frequencies. If our first observation is a photon of frequency $\omega _{mn}$ at time $\tau >0$, then we have $Y_{nm}\left( t\right) $ jumping from value 0 to 1 at $\tau $, with the other channels all registering zero counts. From the SME (\ref{eq:SME}) we therefore have jump
\begin{eqnarray}
\hat{\rho}_{\tau ^{+}}=\hat{\rho}_{\tau ^{-}}+\big\{ |m\rangle \langle m|-%
\hat{\rho}_{\tau ^{-}}\big\} \equiv |m\rangle \langle m|.
\label{eq:jump}
\end{eqnarray}
This just says that the state instantaneously collapses to $|m\rangle $ at time $\tau $. Not surprising as we have just learned that there must have been a transition $n\rightarrow m$ at time $\tau $.

However, we have also learned that the state must have transitioned from the state $|n\rangle $. But where is this in our theory? 

It is \textit{not} the case that $\hat{\rho}_{\tau^-} $ equals $ |n \rangle \langle n | $. Indeed, $ \hat{\rho}_{\tau^-}$ is to be found by solving the ODE (\ref{eq:phat_no_count}) up to time $\tau$ with (some) initial state $\rho_0$: of course, up until $\tau$, all we can say is that no count has been made!

Paradoxically, if we infer a transition from state $|n \rangle$ to $| m \rangle $ at a given time, then we know the state collapses to $\vert m \rangle$ at that time \textit{but it tells us nothing about the state beforehand even though we know it must have been} $|n \rangle$! 

This looks blatantly time-asymmetric. But this is largely due to the fact that we are macroscopic observers causally recording events. Once the transition is observed at time $\tau$, we can make future measurements to check that the state of the system at time $\tau$ was $| m \rangle$. But we do not have the option of going back in time to test whether the state just before $\tau$ was $| n \rangle$. As such, only the information about the state we transition to is important.

Suppose that we have an ancillary system initially prepared in state $\vert \varphi \rangle$ and we measure one of its observables, say $M = \sum_\mu \mu |\mu \rangle \langle \mu |$ where $\{ | \mu \rangle  : \mu \}$ is a complete orthonormal basis for the ancilla. The ancilla observable is to measured at a fixed time $0 < \sigma <T$, where $T$ is the run time for the background continual monitoring as described in the previous sections. The system and ancilla are coupled immediately before $\sigma$ by a unitary of the form $\vert \psi \rangle \otimes | \varphi \rangle \mapsto \sum_\mu \big( \Omega_\mu \rangle
| \psi \big) \otimes | \mu \rangle$ where $\Omega_\mu$ are systems operators which necessarily satisfy $ \sum_\mu \Omega_\mu =\Eye$.

The probability that we have measured $M$ to be eigenvalue $\mu$, conditional on the continuous measurement, is $ q_\mu  = \tilde q _\mu  / \sum_\mu \tilde q_\mu $ where
\begin{eqnarray}
\tilde q _ \mu = \text{tr} \{ \Omega_\mu \hat \rho (\sigma ) \Omega_\mu^\ast \, E (\sigma ) \} .
\label{eq:tilde_q}
\end{eqnarray}
Here $\hat \rho (\sigma )$ is the solution to equation (\ref{eq:SME}) at time $t = \sigma$ with the initial condition $\hat \rho (0) = \rho_0$, and $E(\sigma )$ is an effect value process satisfying the backwards SDE
\begin{eqnarray}  
\overset{\leftarrow}{d} E  ( t)&=& \mathcal{L} \big( E  (   t ) \big) \, dt \label{eq:E_back}\\
& +&
 \sum_{m<n} \bigg(   L^\ast_{mn} E (t  )L_{mn} -  E (   t )\bigg) \, \bigg( \overset{\leftarrow}{d} Y_{mn} (t) -dt \bigg)
,\nonumber 
\end{eqnarray}
with terminal condition $E(T)=\Eye$. (The past-pointing Ito differentials being $ \overset{\leftarrow}{d} X (t) = X( t ) - X (t -dt ) $, for $ dt >0$.)  The derivation of (\ref{eq:E_back}) is that it takes the form of the time-reversed Belavkin-Zakai equation: compare (\ref{eq:BZ_jump}).

The situation we are interested in is where we observe a transition at time $\tau$ after measurement of the ancilla ($\sigma < \tau <T$). We have
\begin{eqnarray}  
\frac{\overset{\leftarrow}{d}E    }{dt} \bigg|_{\text{no count}} &=&   \mathcal{L} \big( E  (   t ) \big) +
 \sum_{m<n} \bigg(   L^\ast_{mn} E (t  )L_{mn} -  E (   t )\bigg)  \nonumber \\
\label{eq:E_no_count}
\end{eqnarray}
valid for times $ t \neq \tau <t $. We propagate (\ref{eq:E_no_count}) back from terminal time $T$ to $\tau^+$. If at the  jump $\tau$ where we observe the transition, say $n \to m$, then (\ref{eq:E_back}) tells us that we have the jump-discontinuity
\begin{eqnarray}
E( \tau^-) &=& L^\ast_{mn} E (\tau^-  )L_{mn} \nonumber \\
&=& \gamma_{mn} \langle m | E( \tau^+ ) m \rangle \, | n \rangle \langle n |.
\label{eq:E_jump}
\end{eqnarray}
We then use (\ref{eq:E_no_count}) to propagate back from $\tau_-$ to $\sigma$.

The discontinuity (\ref{eq:E_jump}) is the resolution of the paradox. $E(\tau^-)$ is proportional to the projection $| n \rangle \langle n|$ on to the pre-transition state.

It is then clear that the probability $q_\mu$ of observing $M= \mu$ (conditional on the transition $n \to m$ at later time $\tau$) for the coupled ancilla depends on the state $\vert n \rangle$.

\section{Conclusion}
We have derived the time-symmetric estimation for indirect measurements made during a continuous monitoring of photons emitted from an open quantum system. The observation of a photon at time $\tau$ reveals that there was a transition $\vert n \rangle \to | m\rangle$ causing the conditioned state to jump from $\hat \rho (\tau^- )$ to the eigenstate $\vert m \rangle$, which ignores that state must have transitioned from $| n \rangle$ at the $\tau$.

Knowledge of the conditioned state $\hat \rho_t$ however is insufficient if we wish to estimate the result of an indirect measurement made at time $\sigma <\tau$, and we need the more complete theory of quantum smoothing \cite{Theory_Past}. Here a critical role is played by (\ref{eq:E_jump}). 
In principle we could run a large number of independent indirect measurements at times $\sigma_1 , \sigma_2 , \dots$ so that we may get arbitrarily close to the random time $\tau $ of the first photon count. 

Suppose that the time 
between an ancilla measurement and the monitored transition is negligible (so that $\tau = \sigma$) then, 
combining (\ref{eq:tilde_q}) and (\ref{eq:E_back}) we have
\begin{eqnarray}
\tilde q _ \mu = \gamma_{mn} \langle n | \Omega_\mu \hat \rho (\tau^- ) \Omega_\mu^\ast \ n \rangle \, \langle m| E (\tau^+ ) |m 
\rangle  ,
\end{eqnarray}
which factors the problem into pre-transition and post transition state terms. 
Moreover the probability is manifestly dependent on $| n \rangle$ and is given by the ratio
\begin{eqnarray}
 q _ \mu = \frac{ \langle n | \Omega_\mu \hat \rho (\tau^- ) \Omega_\mu^\ast \ n \rangle }
{\sum_\mu \langle n | \Omega_\mu \hat \rho (\tau^- ) \Omega_\mu^\ast \ n \rangle }.
\end{eqnarray}
The system has had no time to evolve between the two measurement events and we find that the ratio $q_\mu$ is now wholly independent of the state $\vert m \rangle$ to which the system transitions.

\section*{Appendix A: Derivation of the QSDE}

We now give a microscopic derivation of the QSDE (\ref{eq:U}). The total
Hamiltonian for the system and bath is taken to be $H_{\lambda }=H_{0}+\lambda H_{\text{int}}$ 
where 
\begin{eqnarray*}
H_0 & =& H_{\text{sys}} \otimes \Eye_{\text{bath}}+\Eye_{\text{sys}%
}\otimes \int \omega \left( \mathbf{k}\right) a\left( \mathbf{k}\right)
^{\ast }a\left( \mathbf{k}\right) d^{3}k ,\\
H_{\text{int}} &=& \int \Theta \left( \mathbf{k}\right) \otimes a\left( \mathbf{k%
}\right) ^{\ast }\ d^{3}k+\text{H.c.}
\end{eqnarray*}
Here $a\left( \mathbf{k}\right) $ is the annihilator for a photon of
wave-number $\mathbf{k}$ (we ignore polarizations), $\omega \left( \mathbf{k}%
\right) =c\left| \mathbf{k}\right| $ is the associated energy, and $\Theta
\left( \mathbf{k}\right) $ is an operator on the system space (due to dipole
moments it may have a $\mathbf{k}$ dependence - but the exact form is not
important here.

\begin{figure}[htbp]
	\centering
		\includegraphics[width=0.25\textwidth]{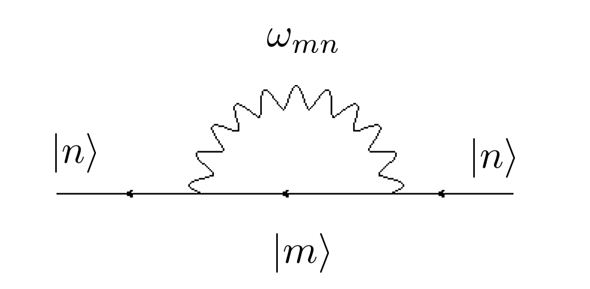}
	\caption{The diagram giving rise to the decay rate $\gamma_{mn}$.}
	\label{fig:transition}
\end{figure}

We shall work in the interaction picture and to this end introduce the
wave-operator $ V\left( t,\lambda \right) =e^{+itH_{0}}e^{-itH_{\lambda }}$.
We next make a Born-Markov approximation. Specifically, this is a weak
coupling limit where we take $\lambda \rightarrow 0$ but rescale time as $%
t/\lambda ^{2}$. The family of unitaries we need is therefore
$ U\left( t,\lambda \right) =V\left( \frac{t}{\lambda ^{2}},t\right) $.
It is not difficult to see that
$ \frac{\partial }{\partial t}U\left( t,\lambda \right) =\sum_{m,n}\left\{
|m\rangle \langle n|\otimes \beta _{mn}\left( t,\lambda \right) ^{\ast }-%
\text{H.c.}\right\} U\left( t,\lambda \right) $,
where we introduce the colored noise fields
$\beta _{mn}\left( t,\lambda \right) =
\frac{-i}{\lambda }\int 
e^{i\left[ \omega \left( \mathbf{k}\right) -\omega _{mn}\right] t/\lambda
^{2}} \, \langle
m|\Theta \left( \mathbf{k}\right) |n\rangle a\left( \mathbf{k}\right) ^{\ast
}\ d^{3}k$.
The processes $b_{mn}\left( t,\lambda
\right) $ converge to quantum white noises as $\lambda \rightarrow 0$. 
Indeed,
$ \left[ \beta _{mn}\left( t, \lambda \right) ^{\ast },\beta
_{m^{\prime }n^{\prime }}\left( s, \lambda \right) \right] $ converges to $\gamma _{nm}\delta
_{nn^{\prime }}\delta _{mm^{\prime }}\delta \left( t-s\right) $
where the $ \gamma _{mn}$ are given by $\int 2\pi \delta \left( \omega \left( \mathbf{k}\right) -\omega
_{mn}\right) \left| \langle m|\Theta \left( \mathbf{k}\right) |n\rangle
\right| ^{2}\ d^{3}k$.
Here we have used the non-degeneracy of the Bohr frequencies together with
the fact that a mismatch between frequencies will result in rapid
oscillation, and so the vanishing of the terms by the Riemann-Lebesgue
Lemma. The $\gamma _{mn}$ are non-zero only for $m<n$ (since $\omega ( \mathbf{k} )  \ge 0$)and give the decay rates associate with the transition $|n\rangle \rightarrow |m\rangle
\rightarrow |n\rangle $ with the $2\pi \delta \left( \omega \left( \mathbf{k}%
\right) -\omega _{mn}\right) $ enforcing the mass-shell condition, see Figure \ref{fig:transition}.

Note that we have dropped Lamb shift terms of the form P.V.$\int \frac{1}{%
\omega \left( \mathbf{k}\right) -\omega _{mn}}\left| \langle m|\Theta \left( 
\mathbf{k}\right) |n\rangle \right| ^{2}\ d^{3}k$ from our considerations as
these may be argued as being negligible, or may be absorbed into the
original system energy eigenvalues (assuming that non-degeneracy still
holds). The limit evolution then describes the  
transitions $|n\rangle $ to a lower energy state $|m\rangle $ and the
creation of a photon in the $\omega _{mn}$ field channel.  

In the limit $\beta _{mn}\left( t\right) $ is approximated by $\sqrt{\gamma
_{mn}}b_{mn}\left( t\right) $ where the $b_{mn}\left( t\right) $ are quantum
input fields. The QSDE (\ref{eq:U}) then corresponds to the open system with
zero-Hamiltonian driven by the $b_{mn}\left( t\right) $ with collapse
operators $L_{mn}=\sqrt{\gamma _{mn}}|m\rangle \langle n|$, for pairs $m<n$.
Note that $K \equiv - \frac{1}{2} \sum_{m<n} L^\ast_{mn} L_{mn}$.


\begin{thebibliography}{99}
\bibitem{martingale}
That is, the conditional expectation of $I_{mn} (s)$ at future time $s>t$ given the observations up to time $t$ is $I_{mn} (t)$.
\end{thebibliography}
\end{document}